# Distinguishing different stackings in WSe$_2$ bilayers grown Using Chemical Vapor Deposition

Aymen Mahmoudi[1], Meryem Bouaziz[1], Davide Romani[1], Marco Pala[1], Aurélien Thieffry[2], Thibault Brulé[2], Julien Chaste[1], Fabrice Oehler[1], Abdelkarim Ouerghi[1]*

[1]Université Paris-Saclay, CNRS, Centre de Nanosciences et de Nanotechnologies, 91120, Palaiseau, France
[2]HORIBA France SAS, Passage Jobin Yvon, Avenue de la Vauve, 91120 Palaiseau, France

The stacking order of two-dimensional transition metal dichalcogenides (TMDs) is attracting tremendous interest as an essential component of van der Waals heterostructures. A common and fast approach to distinguish between the AA′ (2H) and AB (3R) configurations uses the relative edge orientation of each triangular layer (angle θ) from optical images. Here, we highlight that this method alone is not sufficient to fully identify the stacking order. Instead we propose a model and methodology to accurately determine the bilayer configuration of WSe$_2$ using second harmonic generation (SHG) and Raman spectroscopy. We demonstrate that the SHG response of the AB phase (θ = 0°) layers is more intense than the signal from the single layer structure. However, the SHG totally vanishes in the AA′ and AB′ phases (θ = 60° and 0° respectively) of homo-bilayer WSe$_2$. Also, several optical features of homo-bilayer WSe$_2$ are found to depend on the details of the stacking order, with the difference being the clearest in the low frequency (LF) Raman frequencies, as confirmed by DFT simulation. This allows unambiguous, high-throughput, nondestructive identification of stacking order in TMDs, which is not robustly addressed in this emerging research area.

In recent years, interest in van der Waals (vdW) layered materials, including graphite, boron-nitride, and transition metal dichalcogenides (TMDs) have attracted several studies on the structural and electronics properties [1]. These materials display a strong covalent intraplane bonding and a weak vdW interplane bonding, which facilitates their exfoliation into mono-layer or few-layer forms. This further permits to assemble customized stackings made from individual sheets of different chemical composition or crystallographic orientation [2]. Even for a material composed of identical sheets, usually referred to as homo-bilayer or homo-multilayer, the atomic details of stacking order may vary by small translations or rotations, which in turn affects its physical and electronic properties [3].

For TMDs of $MX_2$ formula (M: transition metal; X: chalcogen), various crystalline orderings are referenced, from semiconducting to metallic [3], [4]. With two layers stacked antiparallel, a TMD bilayer with hexagonal stacking (2H) is created and the inversion symmetry is maintained. However, when two layers are stacked parallel, the result is a rhomohedral structure (3R) that breaks out-of-plane mirror symmetry. Broken mirror symmetry leads to charge transfer between layers by hybridization between the occupied and unoccupied states of both layers, generating an out-of-plane electric dipole moment. This makes the 3R phase particularly attractive with ferroelectric properties [5], [6].

The robust control of TMDs crystalline phase is essential for future application, as these semiconducting materials cover band gap window from visible or near-infrared [7], [8] of important technological relevance. For TMD homo-bilayers, there are five high-symmetry stackings predicted, (AA, AB, AA′, A′B, AB′), of which theoretical studies have anticipated that the AA′ (2H) is the most stable, followed by the AB (3R) for $WSe_2$ [9]. Experimentally, CVD growth is reported to produce both AA' (2H) and AB (3R) homo-bilayers [10], [3], [11], [12], [13]. Those two stackings differ by 60° and 0° rotational alignments of the second layer respectively to the first, so optical micrographs may be used as a preliminary sorting tool. Few-layered TMDs can also be fabricated from bottom-up techniques such as molecular beam epitaxy (MBE) [14], [15], [16], or chemical vapor deposition (CVD) [3,17–20]. The CVD process usually produces few tens micrometer-sized monocrystals up to wafer scale size [21–23] which optical quality is now on par with exfoliated material [24]. Moreover, variants of the CVD process can further create graded [25] or abrupt heterostructures [26]. All these materials, being complex heterostructures or simpler homo-bilayers of different stacking order [27] can be probed optically using photoluminescence (PL) and Raman spectroscopy to investigate interlayer coupling and breathing phonon modes [28].

Another informative optical method is second harmonic generation (SHG), which can differentiate between several possible stacking orders and twist angles [29], [30], [3].

In this work, a combination of micro-PL, micro-Raman, and micro-SHG is complemented by theoretical computations based on density functional theory (DFT) to study the stacking order of homo-bilayer $WSe_2$. First, we provide DFT theoretical calculations of the electronic and phononic band structure for each staking. Then our combined experimental and calculated Raman spectroscopies show the correlation between the atomic stacking and the electronic and optical properties of the system. We also have evidence that the second harmonic generation (SHG) intensity of AB phase ($\theta = 0°$, $\theta$ is the relative edge orientation of each triangular layer from optical images) is nearly stronger with respect to the single layer $WSe_2$. However, the SHG signal totally vanishes in the AA′ and in the less considered AB′ phases ($\theta = 60°$ and $0°$ respectively) of homo-bilayer $WSe_2$. Using this combination of optical methods, we sort our $WSe_2$ homo-bilayers between the AA′, AB and, AB′ stacking orders.

The atomic crystal structure of $WSe_2$ consists of planes of covalently bonded W and Se atoms, with a central sublayer of W, sandwiched between two sub-layers of Se [31]. Such planar arrangement exhibits no dangling bond out-of-plane, so homo-bilayer assemblies are held together through vdW forces along the out-of-plane direction [9], [11]. Although AA is the simplest stacking (Figure 1), for which atoms of the same type are superimposed (W/W, Se/Se), it constitutes a very high-energy stacking [10], and hence it is considered energetically unstable [9]. In contrast, the AA′ (2H) order alternates W and Se atoms along the vertical axis (W/Se and vice versa) and has the lowest formation energy. Another stable stacking, although less common, is AB (Figure 1), obtained from AA by translating every other layer by a single bond length. In this case, half of the W atoms are superimposed with half of the Se atoms, and the other half W is located above the empty centers of the lower or upper hexagonal rings. By combining layer rotation and translation, we generate the two remaining stacking orders: A′B and AB′ (Figure 1). In the A′B, there is a complete Se/Se stacking between the two layers, while the W from the top layer lines up with an empty hexagon center from the bottom layer and vice-versa. The (AB′) stacking is the alternate configuration with complete W/W alignment and Se avoidance between the two layers. The total cohesive energies for each homo-bilayer stacking are summarized in Table 1, from theoretical computations (see Methods).

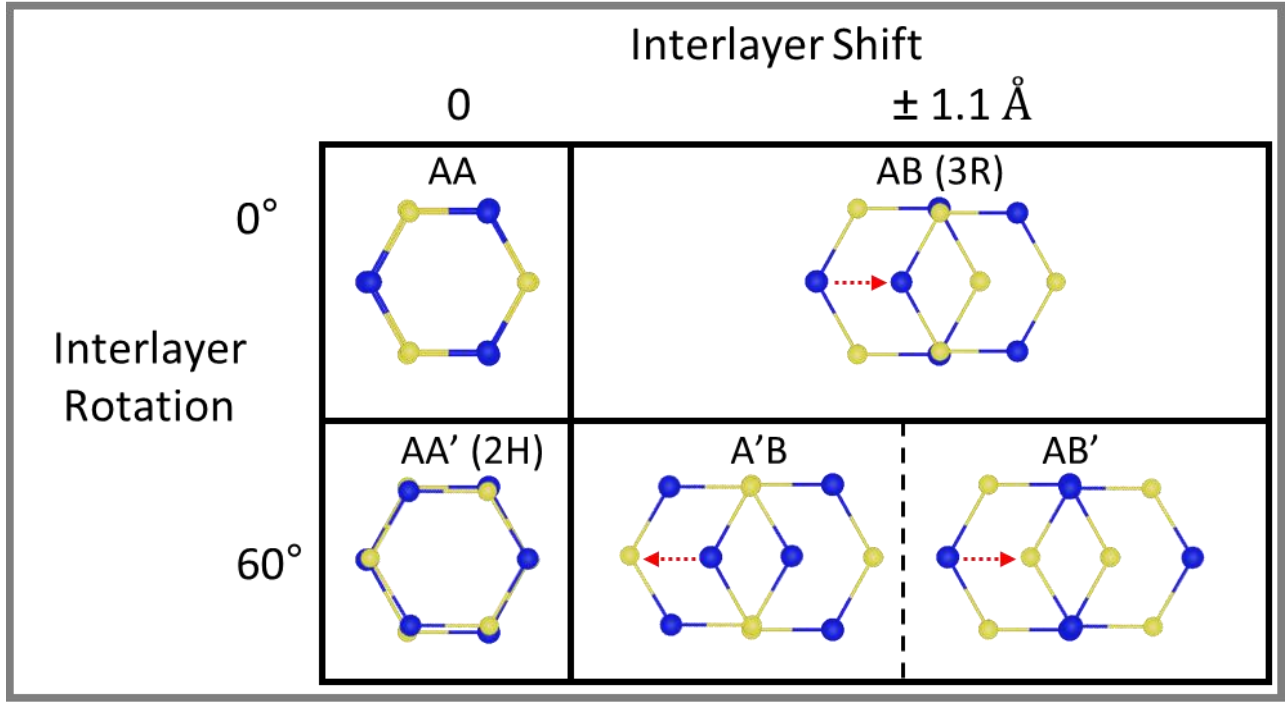

**Figure 1: Crystallography and obtain of various stacking configurations of bilayer WSe$_2$:** Schematic representation describing the structural models of the possible high symmetry of bilayer WSe$_2$ stacking and the geometric relationships between them.

|  | **2H** | **3R** | **AB'** | **A'B** | **AA** |
|---|---|---|---|---|---|
| Relative Energy per unit cell (meV/u.c.) | 0 | 0.4 | 3.4 | 17.4 | 18.6 |
| Plane-Plane distance W-W (Å) | 6.38 | 6.37 | 6.44 | 6.74 | 6.77 |
| Plane-Plane distance Se-Se (Å) | 3.04 | 3.03 | 3.1 | 3.41 | 3.43 |
| Energy Gap (eV) | 1.53 | 1.47 | 1.6 | 1.6 | 1.52 |
| Spin-Orbit Coupling at K (eV) | 0.66 -- 0.00 | 0.56 -- 0.08 | 0.70 -- 0.00 | 0.60 -- 0.00 | 0.57 -- 0.08 |

**Table 1:** The calculated values of physical properties of the bilayer WSe$_2$ structure as a function of stacking order. The relative ground-state energy per cell with respect to the most stable 2H stacking order, the optimized interlayer distance corresponding to tungsten and selenium atoms, band gap energies, spin-orbit coupling, and inter-band splitting values are calculated for the different stacking orders.

We observe that the cohesive energy is strongly correlated to the out-of-plane distance between the two tungsten or selenium atoms from the two layers. Considering the tungsten case for example, the AA′ (2H) and AB (3R) stacking have the smallest interplanar distance of 6.38 Å and 6.39 Å. The W-superposed AB′ (with the avoidance of Se) shows a larger interplanar of 6.44 Å. In contrast, the alignment of Se/Se with W avoidance (A′B) or with W and Se superposition (AA) presents by far the largest interplanar distances of 6.74 Å and 6.77 Å, respectively. Similar calculations for the total cohesive energy have been previously reported for this set of $WSe_2$ homo-bilayers [9]. Therefore, we expect that the unreported AB′ type to be experimentally accessible, in addition to the usual AA′ (2H) and AB (3R) bilayer configurations. If we only consider the two common low-energy configurations, AA′ (2H) and AB (3R) homo-bilayers, simple geometrical considerations using M- or X-zigzag edges allow one to differentiate the two possible stackings, as they differ by a 60° rotation of the upper layer. Such fast analysis can be done directly from optical contrast in conventional micrographs. However, if we include the hypothetically stable AB′ configuration, optical images and edge orientation are not enough to properly differentiate between the possible staking orders.

Here, we performed the CVD growth of wide (several tens of micron) $WSe_2$ homo-bilayers on silica from elemental Se and W oxide (see Methods). By tuning growth duration and the relative position of the substrates, we were able to obtain various homo-bilayer configurations. Figures 2(a, b, c) present an optical micrograph of several $WSe_2$ homo-bilayers. The as-grown $WSe_2$ crystals are easily identified on the silica substrate as triangular shaped flakes. The local thickness, mono-layer (bright) and bi-layer (dark), is also clearly marked by the optical contrast. From the optical images, the alignment of the top layer with respect to the bottom can be rotated by 60° Figure 2(a) or 0° Figure 2(b) and (c). We thus label each sample by this angle "60°", "0° type 1" and "0° type 2". The Figures 2 (d, e, f) present the corresponding SHG intensity map for each crystal using the same intensity scale. Inside a mono- or bi-layer area, the SHG intensity is homogeneous, indicating a uniform and good crystal quality. The SHG intensity is the same for the mono-layer region of each crystal. However, the SHG response from the bilayer area is very different. Quantification from the polar SHG diagram shows that the "0° type 1" structure exhibits a ~4 ratio of SHG intensity between the top layer and bottom layers, which is compatible with the AB (3R) stacking [3]. In contrast, the "60°" and "0° type 2" structures exhibit a clear SHG intensity in the monolayer and a vanishing SHG signal in the bilayer area (Figures 2 (g, h, i)). From the literature, SHG intensity in AA′ (2H) phase $WSe_2$ is known to show an oscillatory trend with even-layered structures having no SHG signal [3]. This originates from the recovering of the inversion symmetry, i.e. centrosymmetric crystal for an even number of layers, so that the resultant SHG fields cancel each other. The similar behavior between the AA′ and AB′ could be explained by the same group symmetry of the two phases, 164 ($D_{3d}$) Whereas AB phase belongs to the 156 ($C_{3v}$) group symmetry. Considering the apparent distinct upper layer rotation 0° and 60° (Figure 2 a and c) and the comparable SHG response (Figure 2 d and f), is a first hint of the existence of an alternative stacking in addition to the well-known set of AB (3R) and AA′ (2H). A likely hypothesis is thus that the "0° type 1" has the AB(3R) configuration, the "60°" is of the AA′ (2H) type and the extra "0° type 2" has the AB′ ordering.

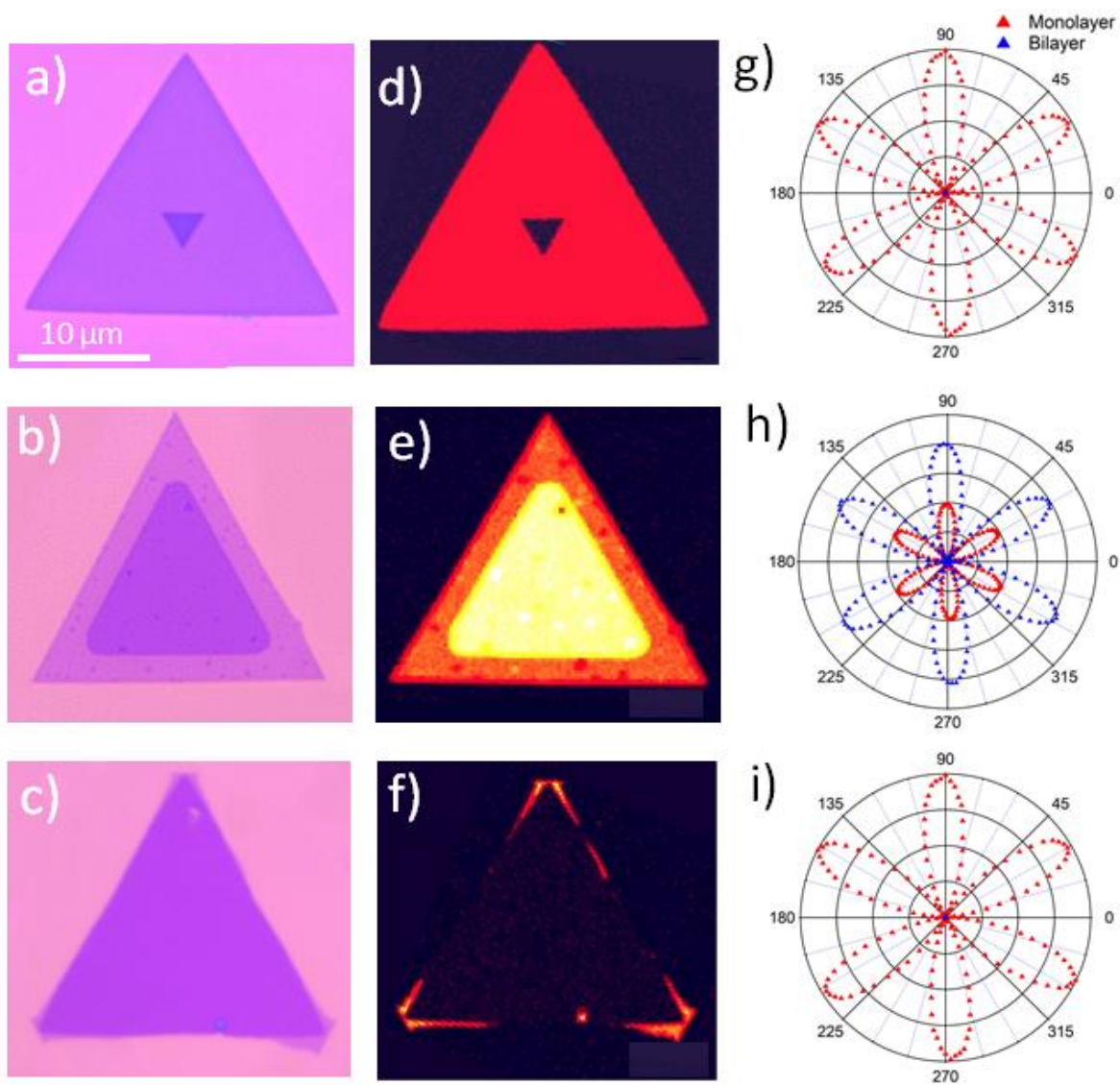

Figure 2: **SHG response of parallel and antiparallel bilayer WSe$_2$ stackings:** (a, b, c) optical and (d, e, f) corresponding SHG maps of the three stacking types. (g, h, i) Corresponding experimental polar SHG intensity diagram taken from the bilayer area of each stacking (1 Ml (red curve), 2 Ml (blue curve).

To confirm the existence and the nature of alternate stacking in the "0° type 2" sample, we now measure its optical properties using micro-PL and micro-Raman. Figure 3(a) compares the room temperature PL spectra obtained from the monolayer and each of the three-bilayer stacking. Compared to the monolayer, the PL intensity is much lower for all bilayers. Such quenching of the PL is predicted for the $WSe_2$ and related to the well-known direct-to-indirect bandgap between 1 and 2ML due to the interlayer electronic coupling [12]. Note that the details of the electronic band structure do depend on the actual ordering of the layer and we present in Supplementary Information Figure S1 the calculated electronic band structure for each ordering. Although the overall electronic dispersions look nearly similar, there are differences for all occupied and unoccupied bands, especially at the top of the valence band near the K point. For instance, the bands at the K point are degenerate for AA′, A′B and AB′, whereas they split into four parabolic branches for AA and AB. Therefore, it is possible to precisely fingerprint each order based on the specifics of its electronic band dispersion, if we exclude the presence of the high energy AA stacking [9]. Besides the overall quenching of the room temperature PL intensity in the bilayer regions, the PL peak energy of the "60°" bilayer is red-shifted by 15 meV compared to the mono-layer reference, that of "0° type 1" by 25 meV, and that of "0° type 2" by 27 meV. Similarly, differences between each staking are also apparent in their PL relative intensity and PL linewidth with respect to the reference monolayer values.

Further insights are obtained by conventional micro-Raman spectroscopy. Figure 3(b) compares the Raman spectra of each polytype near to 250 $cm^{-1}$. The crystals may be distinguished by studying the absolute intensity of the $A_{1g}$ line. The Raman intensity of the "60°" $WSe_2$ bilayer (blue line, Figure 3(b)) is three times larger than the "0° type 1" and the "0° type 2" bilayers (red and green lines respectively, Figure 3(b)) under identical excitation conditions. The relative intensity of the $A_{1g}$ and $E'_{2g}$ can also be used, as "0° type 1" exhibits a clear $E'_{2g}$ signal, while it is nearly absent in the "60°" stacking and the "0° type 2". The corresponding integrated Raman intensity maps are reported in Supplementary Information Figure S4. The variations in the intensity contrast in the Raman intensity map confirm the homogeneity of the different flakes. The presence of both the $A_{1g}$ and $E'_{2g}$ Raman confirms the AB (3R) stacking [11] [10] nature of the flake with "0° type 1". Bilayers with "60°" and "0° type 2" can both be fitted by a single peak at 251 $cm^{-1}$, attributed to the $A_{1g}$, and require further disambiguation. The micro-Raman measurements at different powers or around the optimum optical focus position on both stacking types exhibiting a 0° optical alignment confirm the physical origins of the Raman response and its independence from the excitation power or optical focus (Figure S2 and S3).

In Figure 3(c), we examine the low frequency (LF)range (<50 cm$^{-1}$), where the interlayer in-plane shear ($E_{2g}$) and out-of-plane breathing ($B_{2g}$) modes originate from the relative vibrations between layers. In contrast to the high frequency modes, which are almost unaffected by stacking, these modes are more sensitive to both interlayer coupling and number of layers, enabling a clear identification of layer stacking. We note that the breathing mode near ± 15 cm$^{-1}$ is narrower, with a full-width half-maximum (FWHM) of $1 \pm 0.1$ cm$^{-1}$, compared to the shearing mode at 30 cm$^{-1}$ with a FWHM of $2.6 \pm 0.1$ cm$^{-1}$. The intensity ratio of $E_{2g}/B_{2g}$ is $5.68 \pm 0.03$ for the "60°", which is superior to that of the other staking "0° type 1" ($0.3 \pm 0.03$) and the "0° type 2" ($0.4 \pm 0.03$). In addition to linewidth, we note that the individual peak positions are different for all stackings, which acts as an unambiguous LF Raman fingerprint for each structure. To properly assign each Raman mode to its structure, we show in Figure 3(d) the LF Raman computed from DFT based on relaxed bilayers for the AA′ (2H), AB (3R) and AB′ types of free standing WSe$_2$. We see that each stacking can be assigned to a unique LF Raman fingerprint, with a good match between the theoretical and experimental features. As we move from 2H to 3R stackings in the WSe$_2$ bilayer, the intensity of $B_{2g}$ increases, while the intensity of $E_{2g}$ decreases. This effect was observed for other bilayer TMDs like [32]. In fact, the 3R and AB′ stackings share similar interlayer coupling strength with AA′ stacking, so their intensity changes are caused by the altered atomic arrangements and different electronic environments for each atom. The LF Raman fingerprinting thus determines the "60°" to be the AA′, the "0° type 1" to AB, and the "0° type 2" to be the AB′ stacking. We note that the shift between experimental measurments and calculations is due to some important effects like SiO$_2$ substrate coupling that cannot be properly captured in our modeling of Raman spectra. The intensity estimated by the model predictions is expected to accurately reflect the intrinsic value of a perfect and free standing system, however experimentally the ratio could be significantly altered due to anharmonic coupling of LF modes or selective Raman resonant enhancement of some specific modes.

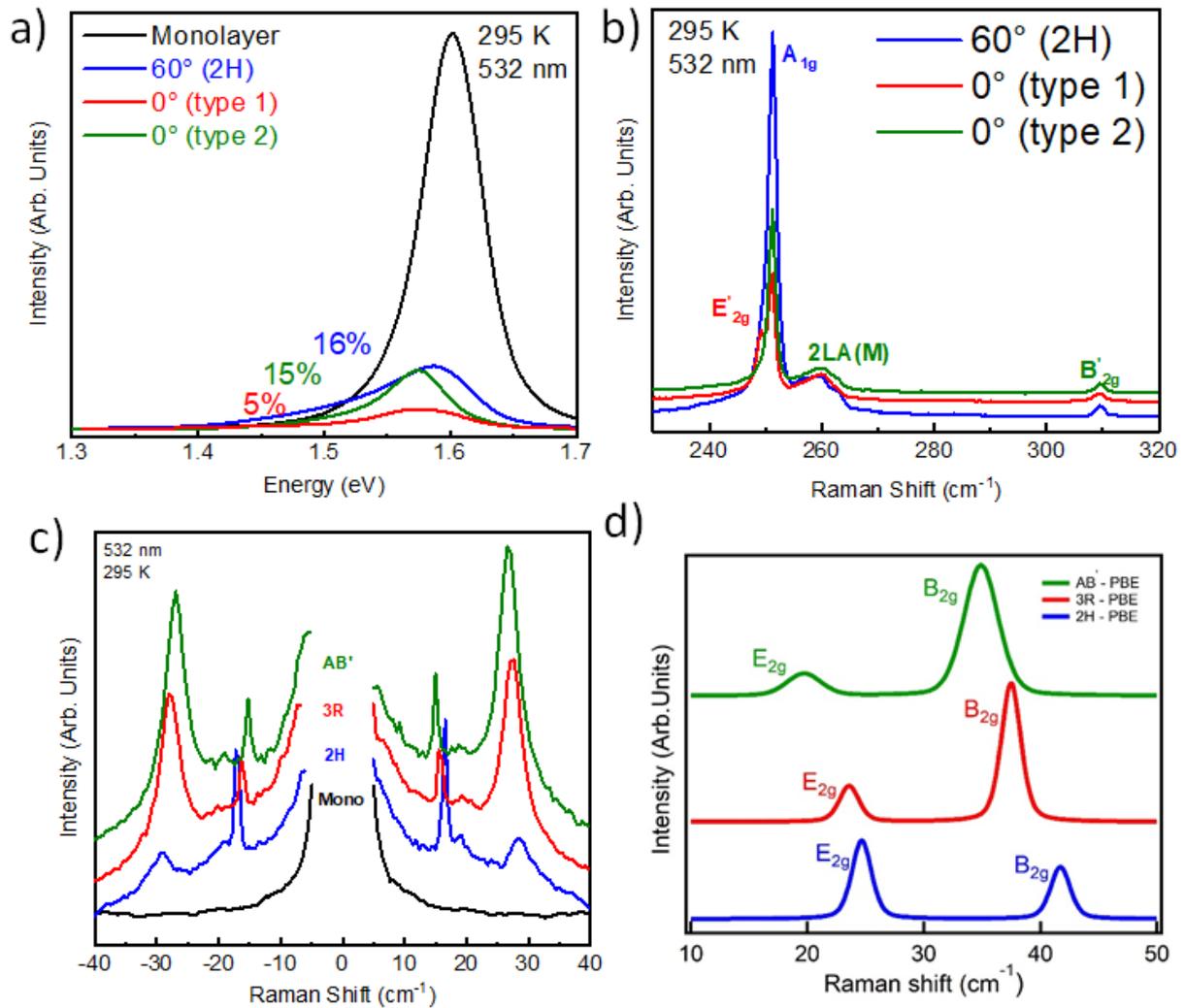

**Figure 3: Comparison of Raman spectra/DFT calculations in the low-frequency range**: (a) and (b) Individual micro-PL and micro-Raman spectra recorded from each WSe$_2$ homo-bilayer stacking 60°-AA′ (blue), 0° type 1 (red), and 0° type 2 (green). (c) Experimental LF Raman spectra for each stacking configurations. (d) Theoretical LF Raman spectra for each stacking. The LF Raman fingerprint assign "60°" to AA′-2H (blue), "0° type 1" to AB-3R (red), and "0° type 2" to AB′ (green).

In Figure 4, we now use this knowledge to study a particular WSe$_2$ crystal with 1, 2- and 3-ML regions, as seen by the optical image Figure 4(a). Optically there is no rotation, i.e. "0° type", between the bottom and intermediate layer. The second layer also appears homogeneous in the optical micrograph, so a reasonable hypothesis would be to consider the complete 2ML region as a standard AB (3R) stacking. Atomic force microscopy (AFM) performed in tapping mode Figure 4(b) confirms that the homogeneous region of the 2ML is of the same height compared to the 1ML and 3ML regions, which agrees with the observed optical contrast. However, the

AFM phase shows the existence of the two domains, with a small phase difference between them. From the AFM topographic data, there is a seamless transition, without a step edge or any other feature, between the two domains at our experimental resolution. However, the micro-Raman intensity map centered around 250 cm$^{-1}$ Figure 4(c) and the micro-PL intensity map Figure 4(d), both indicate the presence of two distinct domains in the 2ML region. Conventional micro-Raman Figure 4(f) shows that the main area of the 2ML region matches the expected AB (3R) micro-Raman features, but that of the lower left area matches the AB′ type. In practice, this turns this particular bi-layer region into a lateral AB-AB′ heterostructure. Figure 4 demonstrates that "nominally" homogeneous 2ML regions from optical micrographs may host alternative stacking, which only appears after detailed PL and Raman investigations.

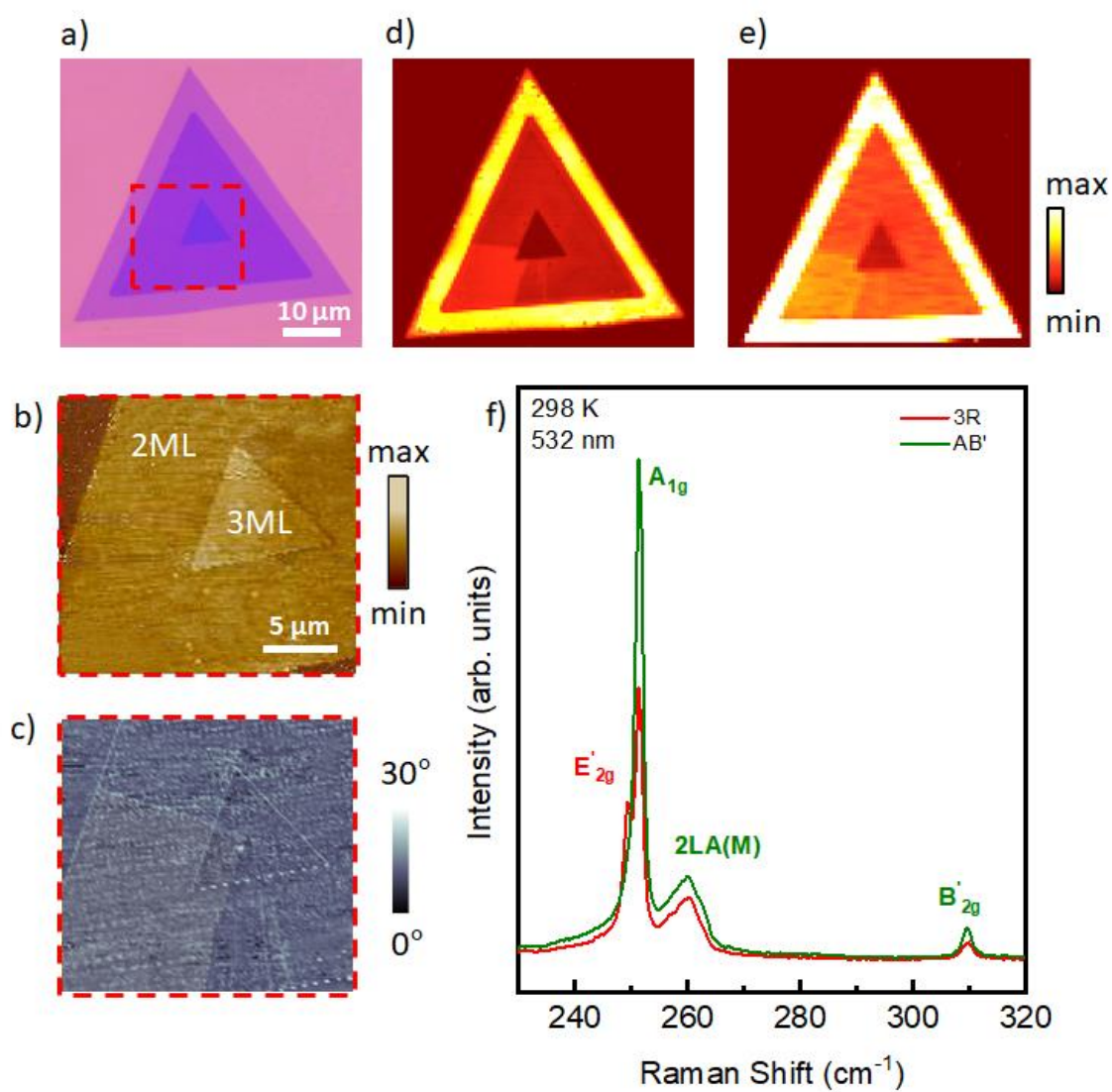

**Figure 4: Lateral heterojunction of AB and AB′ stacking.** (a-b-c) Optical, AFM topography and phase mapping over the bilayer area. The AFM topography image shows no hint of the lateral junction, whereas the AFM phase image reveals a small phase contrast. (d-e) Micro-

Raman and micro-PL intensity maps images of the whole WSe$_2$ crystal. The lateral heterojunction in the lower left bilayer area is clearly visible. (f) Individual micro-Raman spectra obtained from the different zones.

In conclusion, we propose a complete methodology to accurately determine the bilayer configuration of WSe$_2$ by differentiating AA′ (2H), AB (3R) and AB′ based on a combination of optical means, SHG and Raman, which are both non-contact and non-destructive. We confirm that the SHG intensity of AB phase (θ = 0°) homo-bilayers WSe$_2$ is stronger with respect to the single layer WSe$_2$, but that it totally vanishes in the AA′ and AB′ phases (apparent optical edge angle θ = 60 and 0°). Then, we detail both experimentally and theoretically how the micro-Raman is affected by the stacking order in homo-bilayers WSe$_2$, with the difference being the clearest in the low frequency (LF) Raman frequencies, as confirmed by DFT simulation. This allows us to further evidence the presence of "hidden" lateral AB – AB′ heterojunctions in bi-layer regions, whose existence is mostly invisible in conventional optical micrographs. This allows for the unambiguous, high-throughput, and non-destructive identification of stacking order in bi- or few-layer TMDs, which is critically lacking in this emerging research area. We believe that our analysis and methodology will be useful for the growth and fine characterization of more complex TMD heterostructures, supporting the research towards 2D optoelectronic devices.

### III. METHODS:

WSe$_2$ crystals were grown by Chemical Vapor Deposition (CVD) in a customized 4" Nabertherm (RST 120) horizontal oven on SiO$_2$/Si substrates (283 nm thick SiO$_2$ on Si(100) from Nova Wafers). Argon (Ar, 200 sccm) and Hydrogen (H$_2$, 10 sccm) were used as carrier gases. The pressure is 2.5 mbar for the whole growth procedure. The growth duration at high temperature is 30 min, excluding the temperature ramps from and to room temperature (60 min ramp up, 120 min ramp down). During growth, the selenium source is elemental Se (Neyco 99.9%) contained in a quartz crucible heated at 250°C. The tungsten source is H$_2$WO$_4$ (Acros Organics, 99%) powder heated to 850°C, placed upstream to SiO$_2$/Si substrates. The WSe$_2$ homo-bilayers are grown over SiO$_2$/Si substrate (5x2cm) with a preferential location of 3R bilayers downstream, towards the colder part of the substrate. No preferential location could similarly determined for the AB′ stacking. Micro-Raman spectroscopy measurements were conducted at room temperature using a Horiba Labram Raman spectrometer operating at a laser of wavelength λ = 532 nm. The LF Raman spectra were recorded using an ultra-low frequency filter. The laser was focused to a spot size ∼1 μm on the sample, and the spectra were collected using 1800 grooves/mm grating.

First principles calculations for the electronic structure of $WSe_2$ homo-bilayer were performed using plane-wave density functional theory (DFT) as implemented in Quantum ESPRESSO [33]. We have employed full-relativisitic norm-conserving pseudopotentials [34] with an energy cutoff of 60 Ry. The lattice parameter has been set to 3.32 Å, while in the non-periodic z direction we set ~20 Å between the bilayers and employed a Coulomb cut-off strategy to have a proper 2D system [35]. We considered van der Waals interactions via the Grimme-D3 corrections. The atomic positions have been relaxed using the PBE functional and considering spin-orbit interactions with a uniform k-points grid of 12x12x1. The electronic band structure has then been obtained with the HSE06 functional and spin-orbit interaction on a uniform grid of 6x6x1 points for both the k and q vectors (used for the Fock operator), as well as a 120 Ry cutoff energy for the exact exchange operator: the electronic bands have then been obtained via Wannier-interpolation using the Wannier90 code [36] and setting as guess projections d- and p- orbitals for W and Se atoms respectively.

First principles calculations for vibrational properties of $WSe_2$ homo-bilayer were performed using localized Gaussian basis functions as implemented in CRYSTAL17 (CRY) [37]. We have employed the triple-ξ-polarized Gaussian type basis set [38], with real space integration tolerances of 7-7-7-15-30 and an energy tolerance of $10^{-10}$ Ha for the total energy convergence. We employed the PBE exchange-correlation functional together with semi-empirical Grimme's D3 van der Waals corrections. The Brillouin-zone integration is performed on a uniform grid of $18 \times 18 \times 1$ k-points, assuring convergence of both total energy and vibrational frequencies at the center of the Brillouin zone. Freestanding bilayers are modelled with ≈15 Å of vacuum between adjacent layers in the supercell. A lattice parameter of 3.26 Å has been obtained after a geometry relaxation. Polarized Raman spectra were computed as implemented in CRYSTAL17 [37], using experimental conditions (i.e. laser wavelength and temperature) on the relaxed geometry.

**ACKNOWLEDGMENTS**

We acknowledge the financial support by DEEP2D (ANR-22-CE09-0013), 2D-on-Demand (ANR-20-CE09-0026), MixDferro (ANR-21-CE09-0029), Optitaste (ANR-21-CE24-0002), and FastNano (ANR-22-PEXD-0006) projects, as well as the French technological network RENATECH. D.R. acknowledges support from the HPC resources of IDRIS, CINES and TGCC under the allocation 2023-A0140914101 made by GENCI.

**Supporting Information:** See Supplemental Material at [link to be added] for additional Raman spectroscopy measurements on both types 1 and 2 as well as the electronic band structures for the different stacking configurations.

**Competing financial interests:** There are no conflicts to declare.

# Distinguishing different stackings in WSe$_2$ bilayers grown Using Chemical Vapor Deposition

Aymen Mahmoudi[1], Meryem Bouaziz[1], Davide Romani[1], Marco Pala[1], Aurélien Thieffry[2], Thibault Brulé[2], Julien Chaste[1], Fabrice Oehler[1], Abdelkarim Ouerghi[1]*

[1]Université Paris-Saclay, CNRS, Centre de Nanosciences et de Nanotechnologies, 91120, Palaiseau, France

[2]HORIBA France SAS, Passage Jobin Yvon, Avenue de la Vauve, 91120 Palaiseau, France

**Corresponding Author:** abdelkarim.ouerghi@c2n.upsaclay.fr

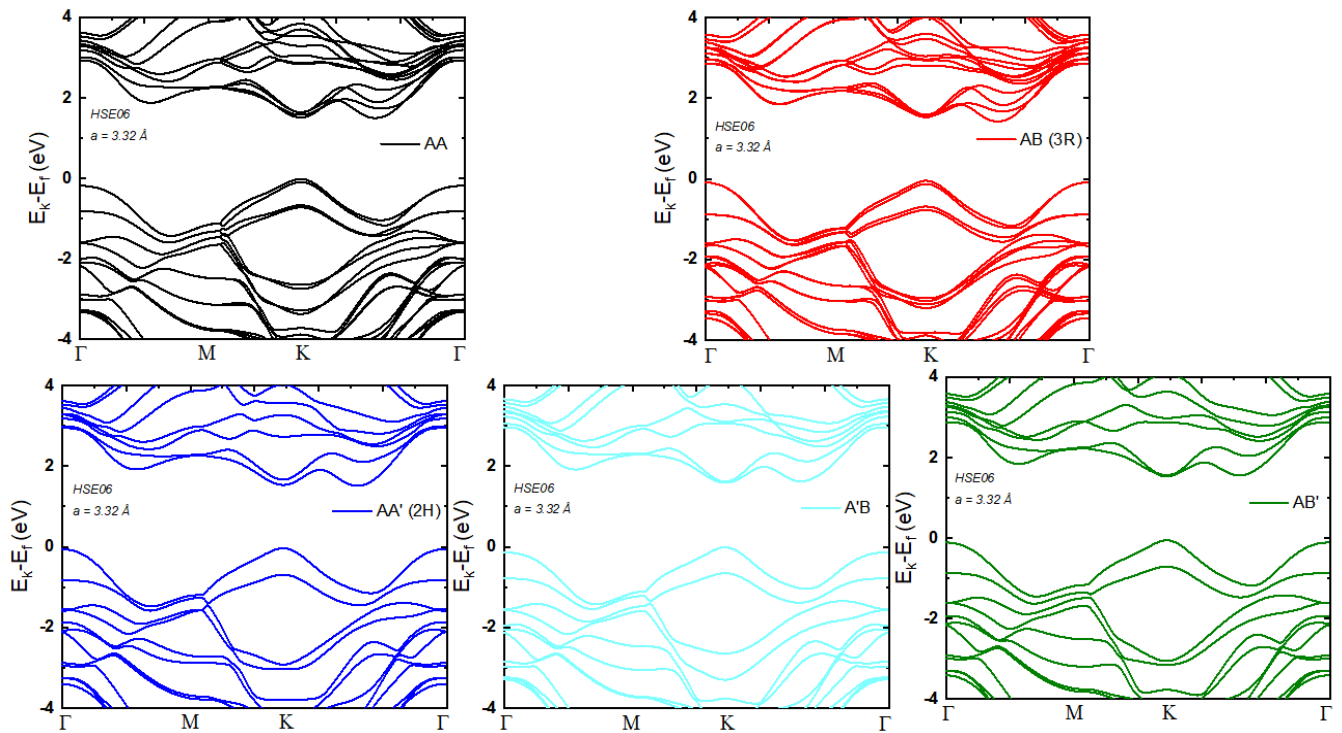

**Figure S1:** Calculated electronic band structure of the five possible high-symmetry stacking orders of bilayer WSe$_2$.

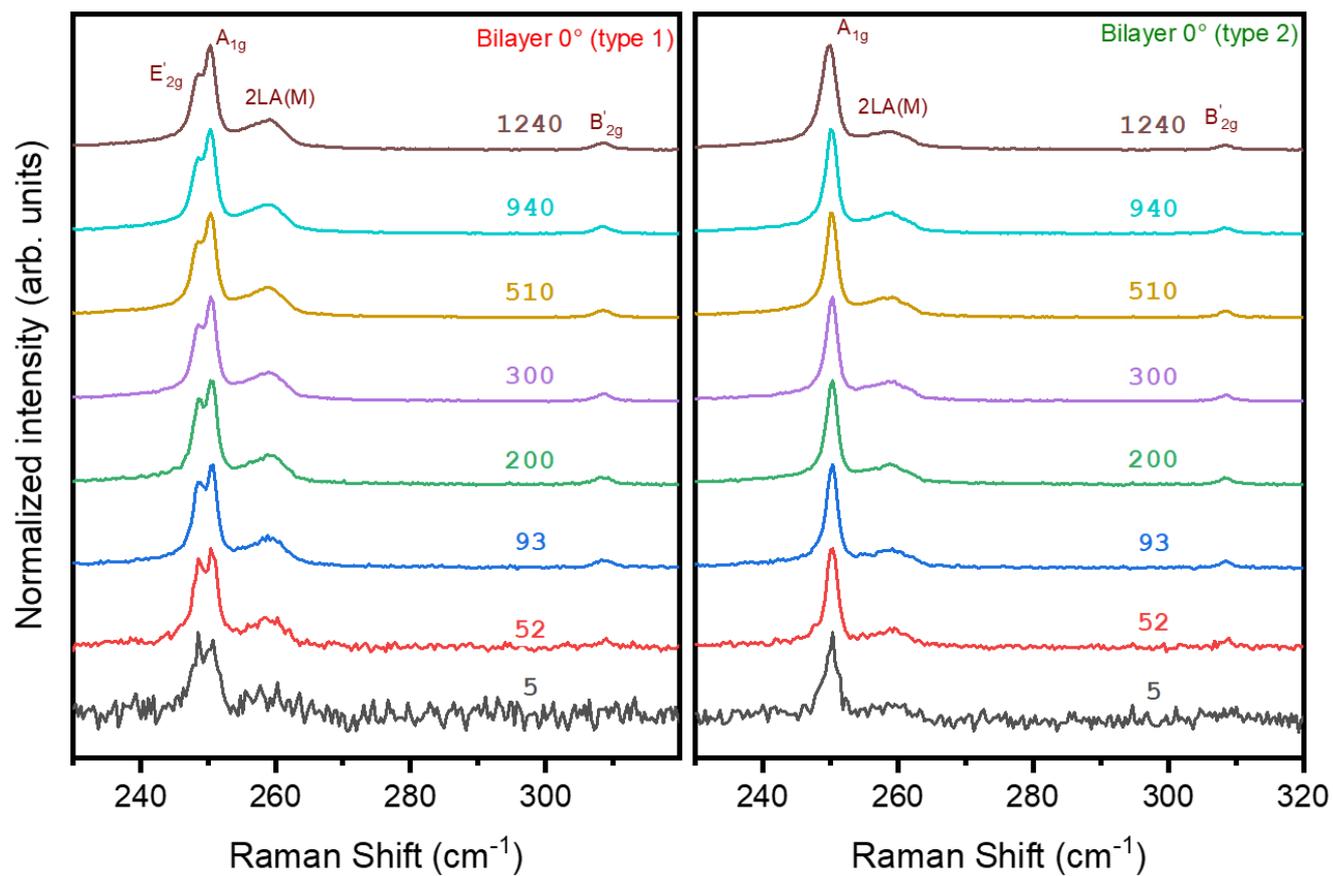

**Figure S2:** Micro-Raman measurements at different powers (5 to 1240 µW) on both stacking types exhibiting a 0° optical alignment. The shape of spectra is well conserved which confirms the physical origins of the Raman response and its independence from the excitation power

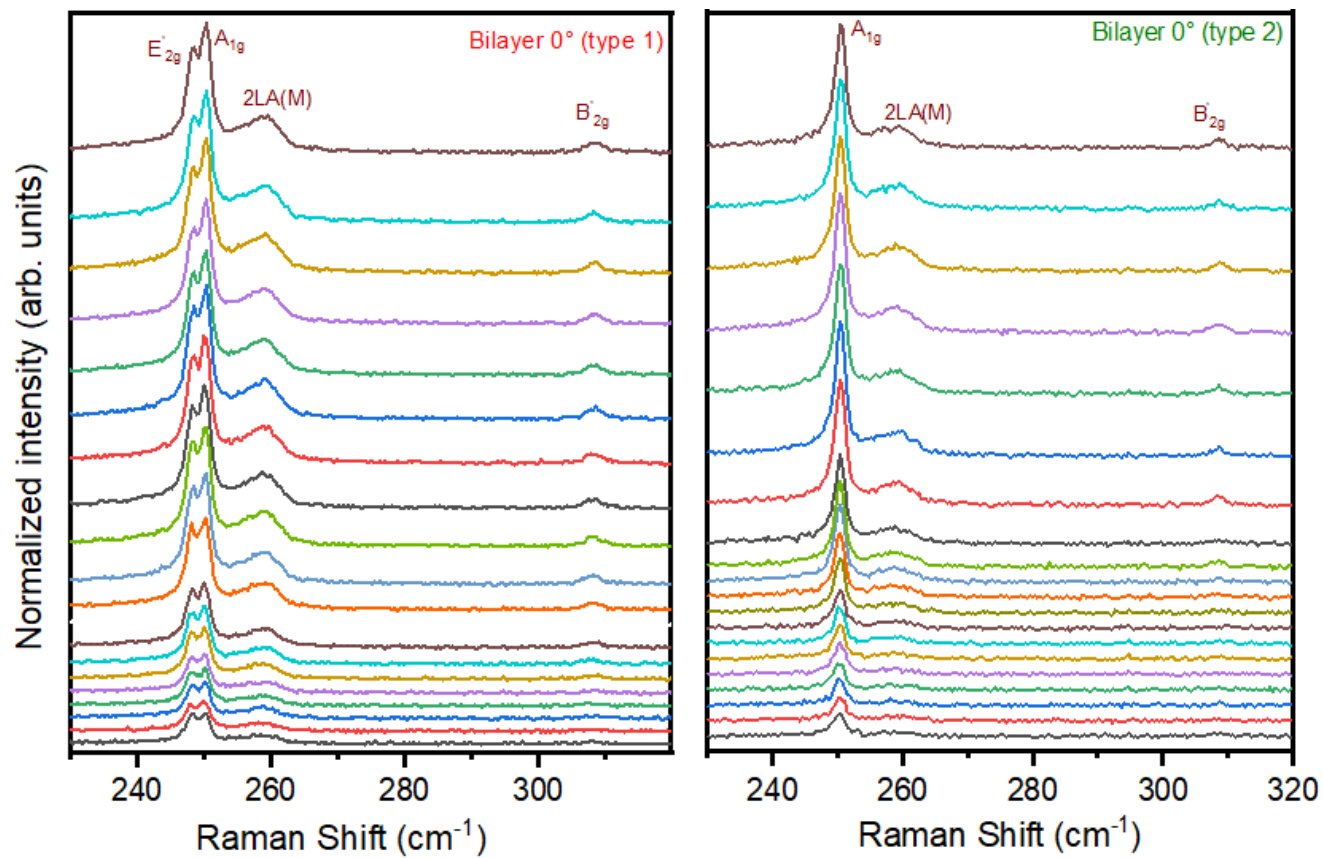

**Figure S3:** Micro-Raman measurements around the optimum optical focus position (from -1 to +1 um) on both stacking types exhibiting a 0° optical alignment. The shape of spectra is well conserved which confirms the physical origins of the Raman response and its independence from the focus position.

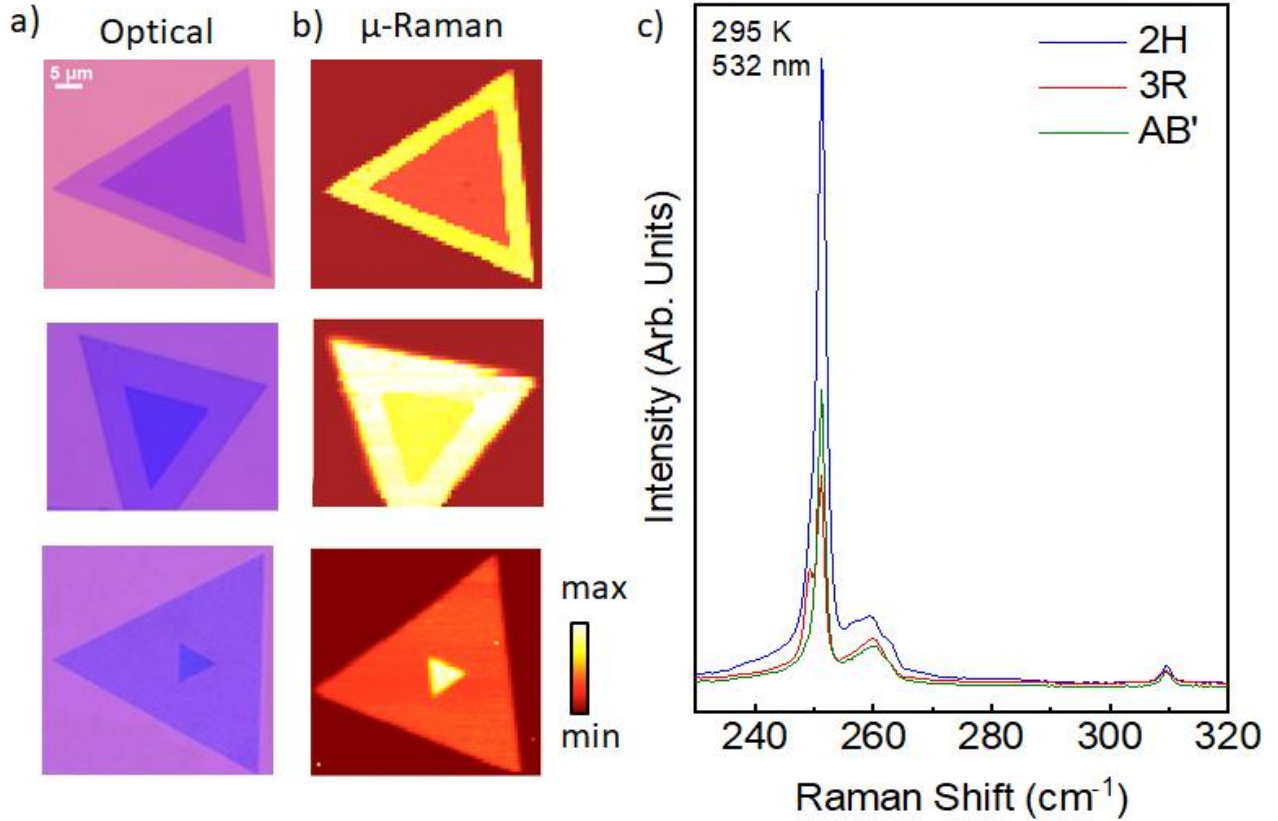

**Figure S4:** Structural properties of three typical homobilayer WSe$_2$/SiO$_2$: (a) optical image, (b) room temperature micro-Raman intensity map. (c) Comparison of individual micro-Raman spectra acquired in the bilayer regions of AB (3R), AB' and AA' (2H) WSe$_2$ crystals (532-nm laser).

Both polytypes exhibit a shear mode in-plane mode and a breathing out-of-plane mode. Our DFT calculations confirm the experimental trend: the frequencies of the first mode are almost the same, while the breathing mode of the 3R phase is ~4 cm-1 less than the shearing mode. This is indeed can be rationalized as a difference in the interlayer bounds: as a matter of fact 3R is less energetically stable and therefore less strongly bounded and then 2H.

Concerning the difference in the Raman response, we follow the discussion outlined in [Luo, X., Lu, X., Cong, C. *et al.* Stacking sequence determines Raman intensities of observed interlayer shear modes in 2D layered materials – A general bond polarizability model. *Sci Rep* **5**, 14565 (2015)] and [Jeremiah van Baren *et al* 2019 *2D Mater.* **6** 025022].

If we fix a vibrational mode k, the associated Raman intensity ($I_k$) is given by the well-known formula:

$$I_k \propto |\vec{e_i} \cdot \boldsymbol{R}_k \cdot \vec{e_j}|^2$$

Where $\vec{e_i}$ and $\vec{e_j}$ are the polarization vectors of the incoming and scattered electromagnetic field, while $\boldsymbol{R}_k$ is the Raman tensor for mode $k$. The latter can be classically computed as:

$$\boldsymbol{R}_k = \frac{\partial \alpha}{\partial Q_k}|_{Q_k=0} \Delta Q_k$$

Where $\alpha$ is the polarizability tensor and $Q_k$ is the normal coordinate of the $k$-th vibrational mode. The derivative is evaluated at equilibrium position.

Raman intensity is therefore closely linked to changes in the system's polarizability caused by atoms' displacements from their equilibrium positions along a specific phonon eigenvector. As a consequence, bonds stretch or compress and significant changes in polarizability due to alterations in the dipole moments (i.e. separation of positive and negative electrical charges between the layers) of these bonds occur. As a matter of fact, while interlayer interactions are largely van der Waals in nature, they also possess a covalent character to some extent. This covalent character allows directional charge accumulation in the bonds, producing angle-dependent dipole moments. Consequently, Raman intensities are influenced by the bond direction, which is here a direct result of the stacking order: without this covalency, completely delocalized interlayer charge distributions would render stacking order insignificant for Raman intensities.

In 3R WSe$_2$ all layers share the same orientation. On the other hand, in 2H WSe$_2$ adjacent layers are oriented oppositely. This difference creates varying bond angles between layers, altering their response to an external electric field (i.e., their polarization) [Nanoscale, 2017, 9, 15340].

Below we give an example for the shearing mode of 2H and 3R bilayer WSe2.

## 2H

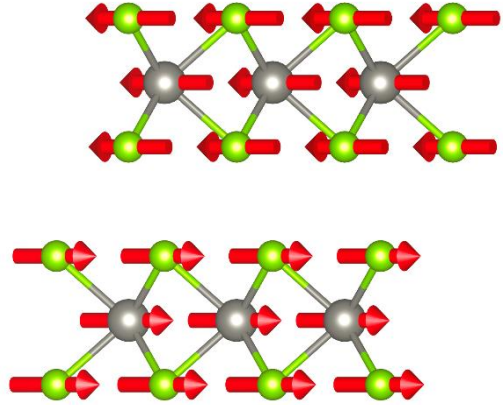

| $Q < 0$ | $Q = 0$ | $Q > 0$ |
|---|---|---|
| 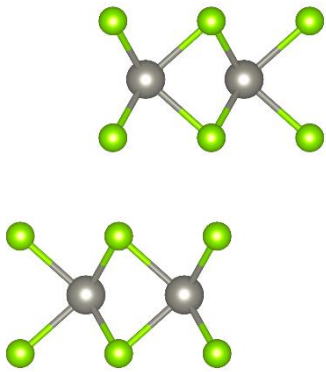 | 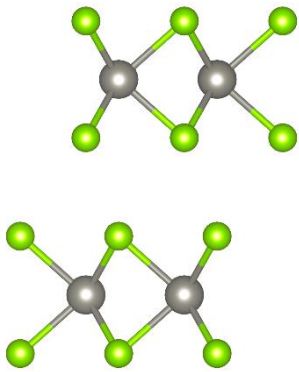 | 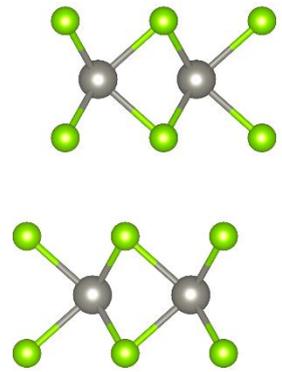 |

## 3R

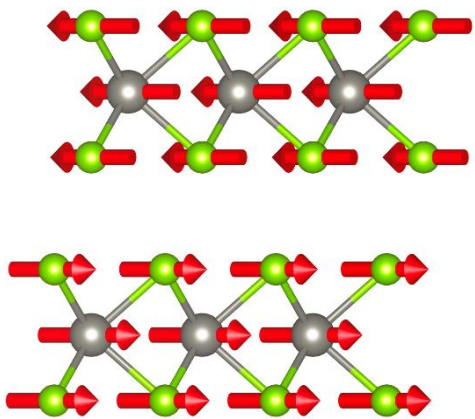

| 3R | | |
|---|---|---|
| $Q < 0$ | $Q = 0$ | $Q > 0$ |

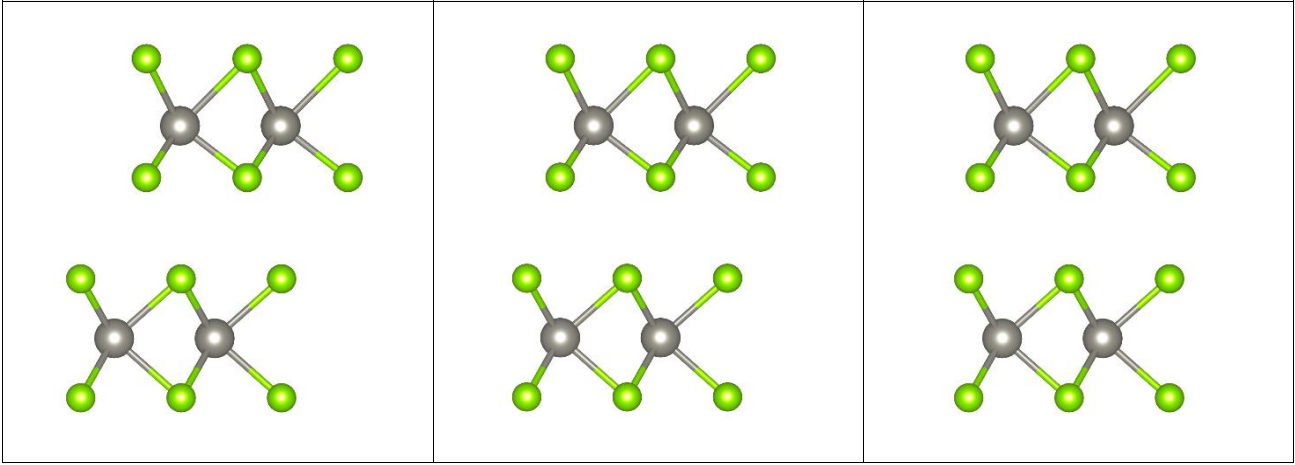